\documentclass[aps,prd,twocolumn,
eqsecnum,showpacs,amsmath,nofootinbib]{revtex4}

\usepackage[dvips]{color,graphicx}
\usepackage{amsfonts,amssymb,theorem,mathrsfs,times}
\newcommand{\qed}{\hbox{\rule[-2pt]{6pt}{6pt}}}

{\theorembodyfont{\upshape}
\newtheorem{Prop}{Proposition}}
{\theorembodyfont{\upshape}
}
{\theorembodyfont{\upshape}
\newtheorem{lm}{Lemma}}
{\theorembodyfont{\upshape}
}
{\theorembodyfont{\upshape}
}
{\theorembodyfont{\upshape}
}

\newcommand{\dalm}{\kern1pt\vbox{\hrule height 0.9pt\hbox{\vrule width
0.9pt\hskip 2.5pt\vbox{\vskip 5.5pt}\hskip 3pt\vrule width 0.3pt}\hrule height
0.3pt}\kern1pt}

\begin{document}


\title{
Black-hole dynamics in BHT massive gravity
}
\author{Hideki Maeda}
\email{hideki-at-cecs.cl}

\address{
Centro de Estudios Cient\'{\i}ficos (CECS), Casilla 1469, Valdivia, Chile
}

\date{\today}

\begin{abstract}
Using an exact Vaidya-type null-dust solution, we study the area and entropy laws for dynamical black holes defined by a future outer trapping horizon in $(2+1)$-dimensional Bergshoeff-Hohm-Townsend (BHT) massive gravity.
We consider the theory admitting a degenerate (anti-)de~Sitter vacuum and pure BHT gravity.
It is shown that, while the area of a black hole decreases by the injection of a null dust with positive energy density in several cases, the Wald-Kodama dynamical entropy always increases.
\end{abstract}



\maketitle

\section{Introduction}
Higher-derivative theories of gravity have been attracting much attention in recent years in the context of quantum gravity.
It is known that such higher-derivative theories typically suffer from the ghost instability.
Nevertheless, higher-curvature corrections in the low-energy limit of string theory provide higher-derivative terms in the field equations in spite of the ghost-free nature of string theory~\cite{bb1996}.
This fact could suggest that there exists a class of well-behaved higher-derivative theories of gravity.

In this context, the forth-derivative theories of gravity formulated by Bergshoeff, Hohm, and Townsend (BHT) in 2+1 dimensions is quite intriguing as a toy model~\cite{bht2009}.
It was shown that this so-called BHT massive gravity is ghost-free around the flat background~\cite{bht2009,bht2009b}, unitary~\cite{no2009}, and renormalizable~\cite{oda2009}.
On the other hand, it was shown that the theory suffers from ghosts around the anti-de~Sitter (AdS) background except for the case with a fine-tuning between the coupling constants~\cite{ls2009}.
(This fine-tuning does not give a degenerate maximally symmetric vacuum but gives the vanishing central charge of the dual theory in the AdS/CFT context.)

Independently, there is another motivation to investigate BHT massive gravity.
The BHT quadratic Lagrangian shares an important property with Lovelock Lagrangian that is proportional to the trace of the resulting field equations.
This is one of the important properties of Lovelock gravity, which is the most general second-order theory of gravity in arbitrary dimensions and contains general relativity as the first-order theory~\cite{lovelock}.
The similarities between BHT massive gravity and the quadratic Lovelock gravity, namely Gauss-Bonnet gravity, have been pointed out by several authors~\cite{or2010,mr2010,sinha2010}.
In this sense, BHT massive gravity may be considered as the simplest higher-derivative gravity and expected to possess some essential properties, which could be inherited to the higher-dimensional counterpart.
(The cubic counterpart of BHT massive gravity in five dimensions was introduced in~\cite{or2010,mr2010}.)

However, because of the complexity of the field equations, it is hard to perform exact analyses or to obtain exact solutions in higher-derivative theories.
In BHT massive gravity, the simplest black hole must be the locally AdS black hole, namely the Ba\~nados-Teitelboim-Zanelli (BTZ) black hole~\cite{btz}.
Interestingly, there is another exact black-hole solution in the theory admitting a single degenerate (A)dS vacuum~\cite{bht2009b,ott2009}.
In this solution, there appears another mysterious parameter as a linear term in the BTZ-type metric function, which causes the slow fall-off to the AdS infinity.
The first law of the black-hole thermodynamics for this black hole was studied by considering the non-trivial contribution of the slow fall-off to the thermodynamical quantities~\cite{gott2009,npy2010}.

However, dynamical aspects of black holes in BHT massive gravity have been totally unclear until now.
The purpose of this paper is to provide an insight for general dynamical properties of black holes in this theory based on an exact model.
The outline of the present paper is as follows.
In section II, we present several exact solutions and their basic properties which are used in the subsequent sections.
In section III, we show the area and entropy laws for a dynamical black hole represented by our new solution.
Our conclusions and future prospects are summarized in section IV.
The Kodama-Wald dynamical entropy is reviewed in Appendix A.
Our basic notations follow \cite{wald}.
The conventions of curvature tensors are 
$[\nabla _\rho ,\nabla_\sigma]V^\mu ={R^\mu }_{\nu\rho\sigma}V^\nu$ 
and $R_{\mu \nu }={R^\rho }_{\mu \rho \nu }$.
The Minkowski metric is taken to be the mostly plus sign, and 
Roman indices run over all spacetime indices.
We adopt the units in which only the gravitational constant $G$ is retained.

\section{Exact solutions in 2+1 BHT massive gravity}
The action $I$ for BHT massive gravity~\cite{bht2009} is given by
\begin{align}
I=& \frac{1}{16\pi G}\int d^3x\sqrt{-g}\biggl(R-2\Lambda-\frac{1}{m^2}K\biggl)+I_{\rm m}, \label{action} \\
K:=&R^{\rho\sigma}R_{\rho\sigma}-\frac{3}{8}R^2, 
\end{align}
where $I_{\rm m}$ is the action for matter.
The gravitational constant $G$ is assumed to be positive.
The field equation is given from the above action as
\begin{align}
&G_{\mu\nu}+\Lambda g_{\mu\nu}-\frac{1}{2m^2}K_{\mu\nu}=8\pi G T_{\mu\nu},\\
&K_{\mu\nu}:=2\nabla^2R_{\mu\nu}-\frac12(\nabla_{\mu}\nabla_{\nu}R+g_{\mu\nu}\nabla^2R)-8R_{\mu\rho}R^\rho_{~~\nu} \nonumber \\
&~~~~~~~~~~~~~~~~~+\frac92 RR_{\mu\nu}+g_{\mu\nu}\biggl(3R^{\rho\sigma}R_{\rho\sigma}-\frac{13}{8}R^2\biggl),
\end{align}
where $K^\mu_{~~\mu}=K$.

\subsection{Maximally symmetric solution}
First we present the locally maximally symmetric vacuum solution in the BTZ-type coordinates: 
\begin{align}
ds^2=&-\biggl(-G\mu+\frac{r^2}{l^2}\biggl)dt^2+\biggl(-G\mu+\frac{r^2}{l^2}\biggl)^{-1}dr^2+r^2d\theta^2,\\
l^{-2}:=&-2m\biggl(m\pm \sqrt{m^2-\Lambda }\biggl),\label{l}
\end{align}
where $\mu$ is a constant.
In this paper, we consider the domain of the coordinate $0\le \theta \le 2\pi$ for $\theta$.
In general there are two distinct maximally symmetric vacua, however the theory admits a degenerate (A)dS vacuum for $\Lambda=m^2$.
For $\mu>0$ and $l^2>0$, this spacetime represents an AdS black hole~\cite{btz}.

\subsection{BTZ-type vacuum solution for $\Lambda=m^2$}
In the theory admitting a degenerate vacuum ($\Lambda=m^2$), there is the following solution~\cite{bht2009b,ott2009}:
\begin{align}
ds^2=&-f(r)dt^2+\frac{dr^2}{f(r)}+r^2d\theta^2, \label{OTT}\\
f(r):=&-G\mu+br+\frac{r^2}{l^2},\label{f}\\
l^{-2}=&-2m^2, \label{l2}
\end{align}
where $\mu$ and $b$ are arbitrary constant.
We call this solution the BHT-Oliva-Tempo-Troncoso (BHT-OTT) solution.
The solution reduces to the BTZ solution for $b= 0$.
There is a curvature singularity at $r=0$ for $b\ne 0$.
The spacetime is asymptotically locally AdS (dS) for $l^2>(<)0$.
The position of a Killing horizon is given by $f(r)=0$, which is solved to give $r=r_{\rm h(\pm)}(>0)$, where 
\begin{align}
r_{\rm h(\pm)}:=&\frac{l^2}{2}\biggl(-b\pm \sqrt{b^2+\frac{4G\mu}{l^2}}\biggl). \label{horizon-d}
\end{align}
The existence and the number of the Killing horizon depending on the parameters are summarized in the Tables 1 and 2.
The black-hole type outer horizon corresponds to $r_{\rm h(+)}$ independent of the sign of $l^2$.
\begin{table}[h]
\begin{center}
\caption{\label{table:th} Number of the Killing horizon in the BHT-OTT spacetime for the asymptotically locally AdS case ($l^2>0$). 
$r_{\rm h(+)}$ and $r_{\rm h(-)}$ are an outer and inner horizons, respectively.
For $\mu<0$ and $b<0$, $-b^2l^2/(4G) \le \mu <0$ is required to have horizons with equality holding for an extremal black hole with a degenerate horizon ($r_{\rm h(+)}=r_{\rm h(-)}$).
}
\begin{tabular}{l@{\qquad}c@{\qquad}c@{\qquad}c}
\hline \hline
  & $\mu>0$ & $\mu=0$ & $\mu<0$   \\\hline
$b>0$ & $r_{\rm h(+)}$ & none & none \\ \hline
$b=0$ & $r_{\rm h(+)}$ & none & none \\ \hline
$b<0$ & $r_{\rm h(+)}$ & $r_{\rm h(+)}(=-bl^2)$  & $r_{\rm h(+)}\ge r_{\rm h(-)}$ \\ 
\hline \hline
\end{tabular}
\end{center}
\end{table} 
\begin{table}[h]
\begin{center}
\caption{\label{table:th2} Number of the Killing horizon in the BHT-OTT spacetime for the asymptotically locally dS case ($l^2<0$). 
$r_{\rm h(+)}$ and $r_{\rm h(-)}$ are an outer and cosmological horizons, respectively.
For $\mu>0$ and $b>0$, $0<\mu \le -b^2l^2/(4G)$ is required to have horizons with equality holding for a degenerate horizon ($r_{\rm h(+)}=r_{\rm h(-)}$).}
\begin{tabular}{l@{\qquad}c@{\qquad}c@{\qquad}c}
\hline \hline
  & $\mu>0$ & $\mu=0$ & $\mu<0$   \\\hline
$b>0$ & $r_{\rm h(-)}\ge r_{\rm h(+)}$ & $r_{\rm h(-)}(=-bl^2)$  & $r_{\rm h(-)}$\\ \hline
$b=0$ & none & none & $r_{\rm h(-)}$ \\ \hline
$b<0$ & none & none & $r_{\rm h(-)}$ \\ 
\hline \hline
\end{tabular}
\end{center}
\end{table} 

Here we review the first law for the BHT-OTT black hole~\cite{gott2009}.
The temperature $T$ and the Wald entropy $S$ for the Killing horizon in the BHT-OTT spacetime are given by 
\begin{align}
T=&\pm \frac{1}{4\pi} \sqrt{b^2+\frac{4G\mu}{l^2}}, \label{T} \\
S=&\pm\frac{\pi l^2}{2G}\sqrt{b^2+\frac{4G\mu}{l^2}},\label{S}
\end{align}
where the sign in the right-hand side corresponds to $r=r_{\rm h(\pm)}$.
It is seen that the Wald entropy of the BHT-OTT black hole is negative in the asymptotically dS case ($l^2<0$).

Using above equations, we obtain
\begin{align}
T\delta S=&\frac{l^2}{16 G}\biggl(2b\delta b+\frac{4G}{l^2}\delta \mu\biggl)
\end{align}
both for $r=r_{\rm h (\pm)}$.
In order to discuss the first law, we have to introduce the mass ${\cal M}$ of a black hole.
In~\cite{gott2009}, mass of the BHT-OTT black hole was defined as
\begin{align}
{\cal M}:=&\frac{\mu}{4}+\frac{l^2b^2}{16G}.\label{mass}
\end{align}  
The above form of the mass was defined to respect the Cardy's formula.
Then the ground state is the extremal configuration in the case of $l^2>0$.
It is easy to see that the first law for the BHT-OTT black hole is satisfied as 
\begin{align}
\delta {\cal M}=T\delta S,\label{1st}
\end{align}
where the parameter $b$ does not appear~\cite{gott2009}.

\subsection{Vaidya-type null-dust solution for $\Lambda=m^2$}
Next we introduce a null dust fluid as a matter field, of which energy-momentum tensor is given by 
\begin{align}
T_{\mu\nu}=\rho l_\mu l_\nu,
\end{align}
where $\rho$ is the energy density and $l^\mu$ is a null vector ($l^\mu l_\mu=0$).
Under the condition $\Lambda=m^2$, there is a following exact solution written in the single null coordinates:
\begin{align}
ds^2=&-g(v,r)dv^2+2dvdr+r^2d\theta^2, \label{vaidya}\\
g(v,r):=&-G\mu+b(v)r+\frac{r^2}{l^2},\\
l^\mu\frac{\partial}{\partial x^\mu}=&-\frac{\partial}{\partial r},\\
\rho(r,v)=&\frac{l^2 bb_{,v}}{16\pi G r},\label{rho}
\end{align}
where a comma denotes the derivative.
We call this solution the Vaidya-BHT solution.
Now $\mu$ is still a constant but $b(v)$ is an arbitrary function of the advanced time $v$.
If $b(v)$ is constant, this solution coincides with the BHT-OTT solution.
Since the Cotton tensor is vanishing in this spacetime, the Vaidya-BHT spacetime is conformally flat.
As a result, it remains a solution even with the gravitational Chern-Simons term in the action, which is put in topologically massive gravity~\cite{tmg}.
We see that the function $b(v)$ is related to the energy density of the null dust.
This fact shows that $b$ surely contributes to the gravitational energy, which is also seen in the mass formula~(\ref{mass}) in the static case.

It must be useful to compare with the corresponding solution in general relativity with a cosmological constant.
The three-dimensional counterpart of the Vaidya-(A)dS solution is given as
\begin{align}
ds^2=&-(-G\mu(v)-\Lambda r^2)dv^2+2dvdr+r^2d\theta^2, \label{vaidya-gr}\\
\rho(r,v)=&\frac{\mu_{,v}}{16\pi r},\label{rho-gr}
\end{align}
where $\mu(v)$ is an arbitrary function~\cite{husain1994}.
We see a sharp difference from the Vaidya-BHT solution in which not $\mu$ but $b$ changes in time and related to the energy density of matter.

\section{The area-entropy laws for the Vaidya-BHT black hole}
In this section, we clarify whether the area and the entropy laws are valid for a dynamical black hole represented by the Vaidya-BHT spacetime.
We define a dynamical black hole by a future outer trapping horizon, which is a local definition of a black hole introduce by Hayward~\cite{hayward1994}.
The future outer trapping horizon expresses the idea that the ingoing null rays should be converging and the outgoing null rays should be instantaneously parallel on the horizon, diverging just outside the horizon and converging just inside.

The position of the trapping horizon in the Vaidya-BHT spacetime is given by $g(v,r)=0$, which is solved to give $r=r_{\rm h(\pm)}(v)$, where $r_{\rm h(\pm)}(v)$ is the same as Eq.~(\ref{horizon-d}) with $b=b(v)$.
Hence, Tables 1 and 2 are still valid for the trapping horizon $r=r_{\rm h(\pm)}(v)$.
Independent of the sign of $l^2$, if $r_{\rm h(+)}\ne r_{\rm h(-)}$, then $r_{\rm h(+)}$ and $r_{\rm h(-)}$ are future outer and future inner trapping horizons, respectively.
If $r_{\rm h(+)}=r_{\rm h(-)}$, we call it a degenerate trapping horizon.

Since Eq.~(\ref{rho}) gives 
\begin{align}
(b^2)_{,v}=&\frac{32\pi G r\rho}{l^2},
\end{align}
the following lemma is shown.
\begin{lm}
\label{lm:b}
In the Vaidya-BHT spacetime under non-negative energy density, $|b|$ is non-decreasing (non-increasing) for $l^2>(<)0$.
\end{lm}
This lemma implies that $b$ cannot be zero from $b\ne 0$ for $l^2>0$, while $b$ cannot change from zero for $l^2<0$.
Namely, a BTZ black hole cannot be realized from a BHT-OTT black hole by the injection of a null dust with positive energy density.

We can also show that the realization of a degenerate trapping horizon depends on the sign of $l^2$.
A degenerate horizon is realized when inside the square-root of Eq.~(\ref{horizon-d}) becomes zero.
Since we obtain 
\begin{align}
\frac{d}{dv}\biggl(b^2+\frac{4G\mu}{l^2}\biggl)=&2bb_{,v}, \nonumber \\
=&\frac{32\pi G r\rho}{l^2}, 
\end{align}
where we used Eq.~(\ref{rho}), the following lemma is shown.
\begin{lm}
\label{lm:degenerate}
In the Vaidya-BHT spacetime with positive energy density, a degenerate horizon can be realized only for $l^2<0$.
\end{lm}

In the following subsections, we consider the following physical setting described by the Vaidya-BHT spacetime: a BHT-OTT black hole with some value of $\mu=\mu_0$ and $b=b_1$ evolutes by the injection of a null fluid with positive energy density to be the one with $\mu=\mu_0$ and $b=b_2$.
The injection of a null dust is assumed to start and cease in such a gentle pace that the spacetime is regular except for the central singularity.
By Lemma~\ref{lm:b}, $b=0$ can be only the initial state for $l^2>0$ or the final state for $l^2<0$.
Since the trapping horizon for $b=0$ with $l^2>0$ is not a black-hole type, we don't consider the case with $b=0$ in the following argument.

\subsection{Area law}
We show the following area law.
\begin{Prop}
\label{area}
({\it Area law.}) 
In the Vaidya-BHT spacetime with positive energy density, the area of the future outer (inner) trapping horizon is increasing (decreasing) for $l^2b<0$.
In the case of $l^2>0$ with $b>0$ ($l^2<0$ with $b<0$), the trapping horizon is future outer (future inner) and its area is decreasing (increasing).
\end{Prop}
{\it Proof}. 
Along the trapping horizon $g(v,r_{\rm h(\pm)})=0$, 
\begin{align}
0=& r_{\rm h(\pm)}b_{,v}dv+\biggl(b+\frac{2r_{\rm h(\pm)}}{l^2}\biggl)dr_{\rm h(\pm)} \label{trapping}
\end{align}
is satisfied. 
Hence we obtain
\begin{align}
\frac{dA_{\rm h(\pm)}}{dv}=&\mp \frac{32\pi^2Gr_{\rm h(\pm)}^2\rho_{\rm h(\pm)}}{l^2 b}\biggl(b^2+\frac{4G\mu}{l^2}\biggl)^{-1/2},\label{key}
\end{align}
where Eqs.~(\ref{horizon-d}) and (\ref{rho}) are used.
$A_{\rm h(\pm)}:=2\pi r_{\rm h(\pm)}$ and $\rho_{\rm h(\pm)}:=\rho(v,r_{\rm h(\pm)}(v))$ are the area of the trapping horizon and the energy density on the trapping horizon, respectively.
The proposition follows from the above equation together with Tables 1 and 2.
\qed

\bigskip

Indeed, the area increasing law is violated for the AdS black hole with $b(v)>0$.
Then, by Lemma~\ref{lm:b}, $b$ is increasing and $r_{\rm h(+)}$ converges to zero for $b \to \infty$ as $r_{\rm h(+)}\simeq G\mu/b$.

By Lemma~\ref{lm:degenerate}, a degenerate horizon is not realized for $l^2>0$.
For $l^2>0$ with $\mu<0$ and $b<0$, we obtain $r_{\rm h(+)}\simeq -bl^2$ and $r_{\rm h(-)}\simeq G\mu/b$ for $b \to -\infty$.
For $l^2<0$ with $\mu>0$ and $b>0$, on the other hand, two horizons can merge to be a degenerate horizon and even disappear with sufficiently large injection of a null dust.

The following signature law is useful to consider the global structure of the spacetime.
The Penrose diagrams representing this process are given in Fig.~1.

\begin{Prop}
\label{signature}
({\it Signature law.}) 
In the Vaidya-BHT spacetime with positive energy density, the future outer (inner) trapping horizon is spacelike (timelike) for $l^2b<0$.
In the case of $l^2>0$ with $b>0$ ($l^2<0$ with $b<0$), the trapping horizon is future outer (inner) and timelike (spacelike).
\end{Prop}
{\it Proof}. 
Using Eq.~(\ref{trapping}) together with Eqs.~(\ref{horizon-d}) and (\ref{rho}), we obtain the line-element on the trapping horizon as
\begin{align}
ds_{\rm h(\pm)}^2=&\mp \frac{l^2b}{8\pi G r_{\rm h(\pm)}^2\rho_{\rm h(\pm)}}\sqrt{b^2+\frac{4G\mu}{l^2}}dr^2+r_{\rm h(\pm)}^2d\theta^2.
\end{align}
The proposition follows from the above equation together with Tables 1 and 2.
\qed


\begin{figure}[htbp]
\begin{center}
\includegraphics[width=0.8\linewidth]{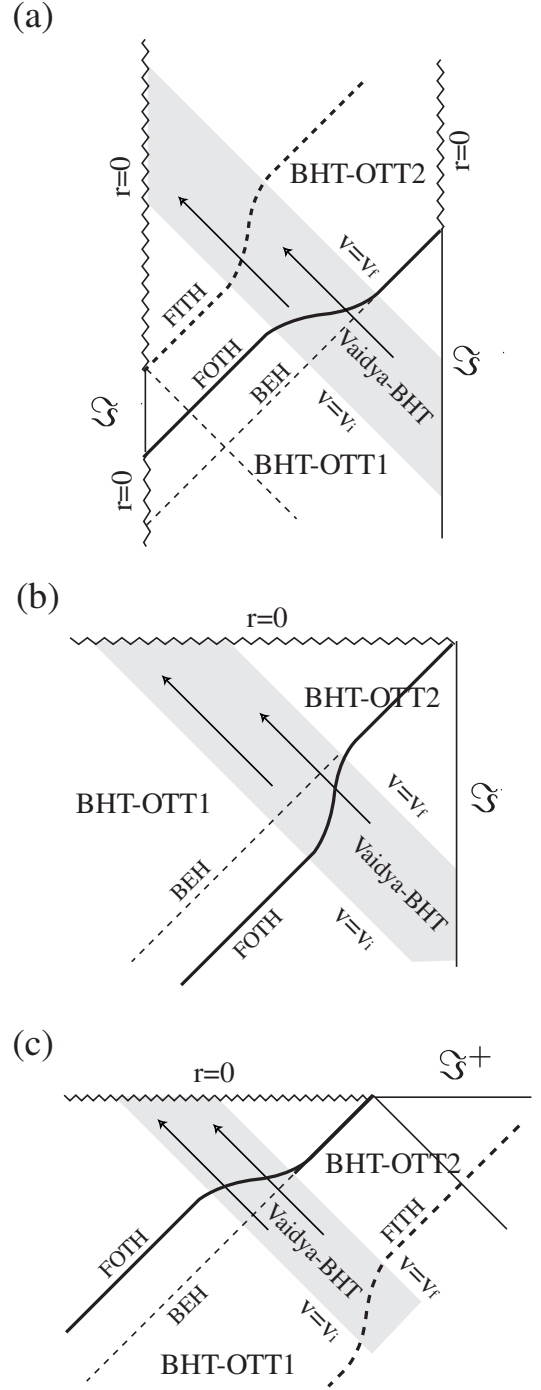}
\caption{\label{Penrose}
A portion of the Penrose diagram representing the transition from 
a BHT-OTT black hole with $\mu=\mu_0$ and $b=b_1$ (BHT-OTT1) 
to another BHT-OTT black hole with $\mu=\mu_0$ and $b=b_2$ (BHT-OTT2)
by an incident null dust fluid with positive energy density in the case of 
(a) asymptotically AdS ($l^2>0$) with $\mu_0<0$ and $b_1<0$, (b) asymptotically AdS ($l^2>0$) with $\mu_0>0$ and $b_1>0$, and (c) asymptotically dS ($l^2<0$) with $b_2>0$ and $0<\mu_0 < -b_2^2l^2/(4G)$.
BEH (a dashed line), FOTH (a thick solid curve), FITH (a thick dashed curve) mean a black-hole event horizon, a future outer trapping horizon, and a future inner trapping horizon, respectively.
Here BHT-OTT1 spacetime for $v \ll v_{\rm i}$ is joined to BHT-OTT2 spacetime 
for $v \gg v_{\rm f}$ by way of the Vaidya-BHT spacetime.
The zigzag line corresponds to a curvature singularity at $r=0$.
}
\end{center}
\end{figure}

\subsection{Entropy law}
We have seen that the area law is violated in the asymptotically AdS case ($l^2>0$) with $b>0$ and $\mu>0$.
Then, how about the entropy law?
To answer this question, we have to define the entropy of a dynamical black hole at the beginning.
In this paper, we adopt the Wald-Kodama dynamical entropy defined on the trapping horizon~\cite{hma1999}.
As shown in Appendix A, the Wald-Kodama entropy of the trapping horizon in the Vaidya-BHT spacetime is given as
\begin{align}
S_{\rm WK}=&\pm\frac{\pi l^2}{2G}\sqrt{b(v)^2+\frac{4G\mu}{l^2}},\label{SWK}
\end{align}
where the sign in the last equation corresponds to $r=r_{\rm h(\pm)}$.
Remarkably, this is the same form as the Wald entropy for the BHT-OTT black hole (\ref{S}).
The following proposition shows that the Wald-Kodama entropy of the black hole is increasing in the process under consideration.

\begin{Prop}
\label{entropy}
({\it Entropy law.}) 
In the Vaidya-BHT spacetime with positive energy density, the Wald-Kodama entropy of the future outer (inner) trapping horizon is increasing (decreasing).
\end{Prop}
{\it Proof}. 
Using Eq.~(\ref{rho}), we obtain
\begin{align}
\frac{dS_{\rm WK}}{dv}=&\pm 8\pi^2 r_{\rm h(\pm)}\rho_{\rm h(\pm)}\biggl(b(v)^2+\frac{4G\mu}{l^2}\biggl)^{-1/2},
\end{align}
where we used Eq.~(\ref{rho}).
The proposition follows from the above equation.
\qed

\bigskip

Here we compare the results with those in general relativity with a cosmological constant.
In the three-dimensional Vaidya-(A)dS spacetime (\ref{vaidya-gr}), the Wald-Kodama entropy of the future outer (future inner) trapping horizon $r_{\rm h}=\sqrt{-G\mu(v)/\Lambda}$ (with $\Lambda<(>)0$ and $\mu>(<)0$) is given by 
\begin{align}
S_{\rm WK}=\frac{\pi}{2G}r_{\rm h}.
\end{align}
In this case, positive energy density requires $\mu_{,v}>0$ and hence the area of the future outer (future inner) trapping horizon and its entropy are both increasing (decreasing).

\subsection{Pure BHT gravity}
Lastly, we present the similar analysis in pure BHT gravity, represented by the action~(\ref{action}) without the Einstein-Hilbert and cosmological constant terms, which has interesting properties shown in~\cite{deser2009}.
Indeed, this theory may be obtained in the (singular) limit of $l^2\to \infty$ in Eq.~(\ref{l}) with $\Lambda=m^2$.
This is because we have $1/l^2=-2m^2=-2\Lambda$ and hence the limit $l^2\to \infty$ gives $\Lambda \to 0$ and $m^2\to 0$, in which the quadratic term dominates in the gravitational action.

In this theory, the BTZ-type static solution is given by
\begin{align}
ds^2=&-(-G\mu+br)dt^2+(-G\mu+br)^{-1}dr^2+r^2d\theta^2, \label{OTT-flat}
\end{align}
where $\mu$ and $b$ are constants~\cite{ott2009}.
The position of the Killing horizon is $r=r_{\rm h}=G\mu/b$ for $b\mu>0$.
For $\mu>0$ and $b>0$, this spacetime represents a locally flat black hole since the Riemann tensor converges to zero for $r\to \infty$.
The temperature and the Wald entropy of the Killing horizon in this spacetime are given by 
\begin{align}
T=&\frac{b}{4\pi}, \label{T-flat} \\
S=&-\frac{\pi b}{4Gm^2}.\label{S-flat}
\end{align}
The Wald entropy of the black hole is positive (negative) for $m^2<(>)0$.

The Vaidya-type null-dust solution is given by
\begin{align}
ds^2=&-(-G\mu+b(v)r)dv^2+2dvdr+r^2d\theta^2, \label{vaidya-flat}\\
\rho(r,v)=&-\frac{bb_{,v}}{32\pi G m^2 r},\label{rho-flat}
\end{align}
where $\mu$ is a constant and $b(v)$ is an arbitrary function.
The Cotton tensor is vanishing in this spacetime.
The position of the trapping horizon is given by $r=r_{\rm h}(v)=G\mu/b(v)$, which is future outer (future inner) for $\mu>(<)0$ and $b(v)>(<)0$.
Since Eq.~(\ref{rho-flat}) gives 
\begin{align}
(b^2)_{,v}=&-64\pi G m^2 r\rho,
\end{align}
$|b|$ is non-increasing (non-decreasing) for $m^2>(<)0$ under non-negative energy density.

The Wald-Kodama entropy of the trapping horizon in this spacetime is given by
\begin{align}
S_{\rm WK}=&-\frac{\pi b(v)}{4Gm^2},
\end{align}
which is the same form as Eq.~(\ref{S-flat}).
We show the following area and entropy laws in this Vaidya-type spacetime.
\begin{Prop}
\label{flat}
({\it Area and entropy laws in pure BHT gravity.}) 
In the pure Vaidya-BHT spacetime (\ref{vaidya-flat}) with positive energy density, the area of the trapping horizon is increasing (decreasing) for $m^2>(<)0$ and the Wald-Kodama dynamical entropy of the future outer (future inner) trapping horizon is increasing (decreasing).
\end{Prop}
{\it Proof}. 
The area of the trapping horizon is given by $A_{\rm h}=2\pi G\mu/b(v)$.
We obtain
\begin{align}
\frac{dA_{\rm h}}{dv}=&\frac{64\pi^2 G^3\mu^2 m^2\rho_{\rm h}}{b^4}
\end{align}
and 
\begin{align}
\frac{dS_{\rm WK}}{dv}=&\frac{8\pi^2 G \mu\rho_{\rm h}}{b^2},
\end{align}
where we used Eq.~(\ref{rho-flat}).
The proposition follows from the above equations together with the fact that future outer (future inner) trapping horizon corresponds to $\mu>(<)0$.
\qed

\section{Summary and discussions}
In this paper, we have discussed the black-hole dynamics in BHT massive gravity.
We have considered the theory with $\Lambda=m^2$, which admits the theory to have a degenerate (A)dS vacuum, and pure BHT gravity.
We have obtained an exact Vaidya-type null-dust solution representing a transition of a black hole to another by the injection of a null dust fluid with positive energy density and clarified the area and entropy laws.

In this model, the area law for a dynamical black hole defined by a future outer trapping horizon is violated for $m^2<0$ in the asymptotically locally AdS (with $b>0$ and $\mu>0$) and flat cases.
We have shown that, in spite of the violation of the area law, the Wald-Kodama dynamical entropy of a black hole is increasing independent of the parameters.
This result surely suggests that the entropy law is more fundamental than the area law in black-hole physics.
Actually, this result is quite similar to the case in Einstein-Gauss-Bonnet gravity~\cite{nm2008,maeda2010}.

The area law for a black hole is satisfied for $m^2>0$.
However, it is difficult to claim by the present analysis that the theory with $m^2>0$ has nicer properties than with $m^2<0$.
The first reason is that the Wald-Kodama entropy of a black hole is always negative for $m^2>0$.
Although we still don't know the consequence of the negative entropy of a black hole, it could be interpreted as a certain kind of instability stemming from the absence of a microscopic theory describing the black hole.
The second reason is that the area {\it decreasing} law for a future {\it inner} trapping horizon is actually violated for $m^2>0$ in the asymptotically locally dS (with $b<0$) and flat cases.
These area laws are related to the attractive nature of gravity, which is characterized by the null convergence condition in the Raychaudhuri equation.
The null convergence condition is equivalent to the null energy condition in general relativity and therefore the violation of the area law under the null energy condition suggests that the BHT quadratic term acts as an exotic matter.

Our results suggest that the entropy law is still valid in BHT massive gravity in more general cases.
A definitely important future task is to clarify the generic properties of a dynamical black hole under the null or dominant energy condition without specifying a matter field.
This is a problem worth challenging in order to understand the nature of higher-derivative theories of gravity. 
Even in the symmetric spacetime considered in this paper, it is difficult to evaluate the higher derivative terms in the field equations.
In Einstein-Gauss-Bonnet gravity, the generalized Misner-Sharp quasi-local mass played a key role to show them, which is the first integral of the field equations~\cite{maeda2006b,mn2008}.
This quantity satisfies the unified first law and can be constructed as a locally conserved quantity associated with the Kodama observer.
However, in BHT massive gravity, the integrability condition for the unified first law is not satisfied and hence such a quantity cannot be constructed in the same way.
(The situation is similar in $f(R)$ gravity~\cite{ccho2009}.)
A new method is highly required in order to control the higher derivative terms.

\acknowledgments 
The author thanks J.~Oliva, S.~Ray, and R.~Troncoso for useful comments. 
This work has been partially funded by the Fondecyt grant 1100328 and by the Conicyt grant "Southern
Theoretical Physics Laboratory" ACT-91. 
The Centro de Estudios Cient\'{\i}ficos (CECS) is funded by the Chilean Government through the Centers of Excellence Base Financing Program of Conicyt.

\appendix

\section{Wald-Kodama dynamical entropy of the Vaidya-BHT black hole}
The Wald entropy $S$ in arbitrary $n(\ge 3)$ dimensions is defined for a Killing horizon generated by a Killing vector $\xi^\mu$.
Here we focus on the gravitation theory of which action $I$ is given by
\begin{align}
I=&\int\varepsilon_{\mu_1\mu_2\cdots \mu_{n}}{\cal L}(g^{\mu\nu},R_{\mu\nu\rho\sigma})dx^{\mu_1}\wedge \cdots \wedge dx^{\mu_n},
\end{align}
where $\varepsilon_{\mu_1\mu_2\cdots \mu_{n}}dx^{\mu_1}\wedge \cdots \wedge dx^{\mu_n}$ is the volume $n$-form.
The Wald entropy is defined by the following integral performed on $(n-2)$-dimensional spacelike bifurcation surface $\Sigma$~\cite{wald1993,iyerwald1994,jkm1994,km1998}:
\begin{align}
S:=&\frac{2\pi}{\kappa} \oint {\bf Q},\label{WaldS}
\end{align}
where $\kappa$ is the surface gravity of the Killing horizon, defined by $\xi^\nu \nabla_\nu \xi_\mu =\kappa \xi_\mu$ evaluated on the Killing horizon.
${\bf Q}$ is the Noether charge $(n-2)$-form defined by 
\begin{align}
{\bf Q}:=&\frac12 \varepsilon_{\mu\nu\alpha_1\cdots \alpha_{n-2}}Q^{\mu\nu}dx^{\alpha_1}\wedge \cdots \wedge dx^{\alpha_{n-2}},\\
Q^{\mu\nu}:=&-2X^{\mu\nu\rho\sigma}\nabla_\rho \xi_\sigma+4\xi_\sigma \nabla_\rho X^{\mu\nu\rho\sigma},\label{Q} \\
X^{\mu\nu\rho\sigma}:=&\frac{\partial {\cal L}}{\partial R_{\mu\nu\rho\sigma}}.
\end{align}
It is noted that $\xi^\mu=0$ is satisfied on $\Sigma$, and hence the second term in Eq.~(\ref{Q}) does not make any contribution to the Wald entropy.
It is noted that $\nabla_\rho X^{\mu\nu\rho\sigma}\equiv 0$ is satisfied for Lovelock gravity. 
(See section 16.4 in~\cite{pad}.) 

The Wald-Kodama entropy is given by Eq.~(\ref{WaldS}) with $\xi^\mu$ replaced by a Kodama vector $K^\mu$, integrated over the trapping horizon~\cite{hma1999}.
$\kappa$ in Eq.~(\ref{WaldS}) is now the dynamical surface gravity of the trapping horizon, defined by $K^\nu \nabla_{[\nu} K_{\mu]} =\kappa K_\mu$ evaluated on the trapping horizon.
A Kodama vector~\cite{kodama1980} is defined on the cross-product manifold ${\cal M}^n\approx M^2\times K^{n-2}$, where $(M^2,g_{ab})$ is a two-dimensional Lorentzian spacetime and $(K^{n-2},\gamma_{ij})$ is an $(n-2)$-dimensional Riemannian space, by
\begin{align}
K^{\mu}:=&-\varepsilon^{\mu\nu}\nabla_\nu r.
\end{align}
The most general metric on this product spacetime $({\cal M}^n, g_{\mu\nu})$ is written as
\begin{align}
ds^2=g_{ab}(y)dy^a dy^b +r(y)^2\gamma_{ij}(z)dz^idz^j.
\end{align}
Here $\varepsilon_{\mu \nu}=\varepsilon_{ab}(d x^a)_{\mu}(d x^b)_{\nu}$, where $\varepsilon_{ab}$ is a volume element on $(M^2, g_{ab})$. 

For BHT massive gravity, we obtain 
\begin{align}
\frac{\partial R}{\partial R_{\mu\nu\rho\sigma}}=&g^{\nu[\sigma}g^{\rho]\mu},\\
\frac{\partial (R_{\alpha\beta}R^{\alpha\beta})}{\partial R_{\mu\nu\rho\sigma}}=&2g^{\alpha[\nu}g^{\mu][\rho}g^{\sigma]\beta}R_{\alpha\beta}
\end{align}
and hence
\begin{align}
X^{\mu\nu\rho\sigma}=&\frac{1}{16\pi G}\biggl[\biggl(1+\frac{3}{4m^2}R\biggl)g^{\nu[\sigma}g^{\rho]\mu} \nonumber \\
&-\frac{2}{m^2}g^{\alpha[\nu}g^{\mu][\rho}g^{\sigma]\beta}R_{\alpha\beta}\biggl].
\end{align}

In the Vaidya-BHT spacetime (\ref{vaidya}), we have $\varepsilon_{vr}=1$ and $\varepsilon^{vr}=-1$ and therefore the non-zero component of $K^\mu$ is $K^v=-\varepsilon^{vr}=1$.
The dynamical surface gravity on the trapping horizon is given by $\kappa=(\partial g(v,r)/\partial r)/2|_{r=r_{\rm h (\pm)}}$.
The non-zero component of $Q^{\mu\nu}$ is given by
\begin{align}
Q^{vr}=-Q^{rv}=&\frac{(b(v)l^2+2r)^2}{16\pi Gl^2r}.
\end{align}
Finally, the Wald-Kodama entropy is calculated to give
\begin{align}
S_{\rm WK}=&\frac{\pi(2r+b(v)l^2)}{2 G} \biggl|_{g(v,r)=0}, \nonumber \\
=&\pm\frac{\pi l^2}{2G}\sqrt{b(v)^2+\frac{4G\mu}{l^2}},
\end{align}
which is the same form as the Wald entropy for the BHT-OTT black hole (\ref{S}).

It is noted that $\nabla_\rho X^{\mu\nu\rho\sigma}$ is non-vanishing in general for BHT massive gravity.
Nevertheless, $\nabla_\rho X^{\mu\nu\rho\sigma}= 0$ is satisfied for the Vaidya-BHT spacetime.
This is the reason why the Wald-Kodama entropy coincides with the Wald entropy in the static case.

For the pure Vaidya-BHT spacetime (\ref{vaidya-flat}), the non-zero component of $Q^{\mu\nu}$ is given by
\begin{align}
Q^{vr}=-Q^{rv}=&-\frac{b^2}{32\pi Gm^2r},
\end{align}
while the dynamical surface gravity is given by $\kappa=b/2$.
Then, the Wald-Kodama entropy is calculated to give
\begin{align}
S_{\rm WK}=&-\frac{\pi b(v)}{4Gm^2},
\end{align}
which is the same form as Eq.~(\ref{S-flat}) since $\nabla_\rho X^{\mu\nu\rho\sigma}= 0$ is also satisfied for this spacetime.


\end{document}